\documentclass[prl,twocolumn,aps,showpacs,preprintnumbers,amsmath,amssymb,floatfix]{revtex4}
\usepackage{graphicx}
\usepackage{epsfig}
\usepackage{color}

\newcommand{\ark}{{\rm\bf k}}
\newcommand{\khat}{\hat{K}}
\newcommand{\hhat}{\hat{H}}
\newcommand{\dhat}{\hat{D}}
\newcommand{\Nhat}{\hat{N}}
\newcommand{\curO}{{\cal {\hat O}}}
\newcommand{\chat}{\hat{c}}

\newcommand{\Hh}{\hat{H}}
\newcommand{\nhat}{\hat{n}}
\newcommand{\be}{\begin{equation}}
\newcommand{\ee}{\end{equation}}
\newcommand{\bea}{\begin{eqnarray}}
\newcommand{\eea}{\end{eqnarray}}
\newcommand{\kb}{k_{\rm B}}
\begin{document}
\title{The response to dynamical modulation of the optical lattice for fermions in the Hubbard model}
\author{Zhaoxin Xu$^1$,
Simone Chiesa$^{2,3}$,
Shuxiang Yang$^1$, 
Shi-Quan Su$^1$,  
Daniel E. Sheehy$^1$, 
Juana Moreno$^1$,
Richard T. Scalettar$^2$, and
Mark Jarrell$^1$}

\affiliation{$^1$Department of Physics $\&$ Astronomy, 
Louisiana State University, Baton Rouge, LA 70803, USA
\\
$^2$Physics Department, University of California, Davis, CA 95616, USA\\
$^3$Physics Department, University of Tennessee, Knoxville, TN 37996, USA}
\date{\today}
 
\begin{abstract}
Fermionic atoms in a periodic optical lattice provide a realization of
the single-band Hubbard model.  Using Quantum Monte Carlo simulations
along with the Maximum Entropy Method, we evaluate the
effect of a time-dependent perturbative modulation of the optical
lattice amplitude on atomic correlations, revealed in the fraction of
doubly-occupied sites.  Our treatment extends previous approaches which
neglected the time dependence of the on-site interaction, and shows that
this term changes the results in a quantitatively significant way.
The effect of modulation depends strongly on the filling-- the response
of the double occupation is significantly different in the half-filled
Mott insulator from the doped Fermi liquid region.
\end{abstract} 

\maketitle

A number of key properties of strongly correlated electron
systems appear to be well described by simplified tight-binding
Hamiltonians.  For example, the square lattice Hubbard model, with one
particle per site, is known to possess the long range antiferromagnetic
order manifest in the parent compounds of high temperature
superconductors, whose CuO$_2$ sheets have square arrays of copper atoms
with one hole per $3d$ shell.  There are many analytic and numerical
clues that suggest the doped Hubbard model might also possess the
$d$-wave superconducting phase exhibited by the cuprates, as well as
other non-trivial properties including stripes and pseudogap physics
\cite{scalapino94}.  If this could be demonstrated rigorously, it would
provide important insight into the mechanism of superconductivity in
these materials.

Ultracold atomic systems  
offer an opportunity for closer connection between experiments and
calculations for such model Hamiltonians.  At present, experiments on
fermionic atoms are exploring temperatures $T$ which are of the order of
the hopping integral $J_0$, probing correlations such as double
occupancy, $D$, and short range spin order that develops at that
temperature scale.  In particular, the evolution of $D$ with the ratio
of interaction strength $U$ to hopping $J_0$ has been shown to indicate
the presence of a Mott metal-insulator transition
\cite{jordens08,schneider08}.  The presence of a Mott gap in the
excitation spectrum has also been inferred through peaks in $D$ which
arise through a dynamic modulation of the optical lattice depth $V$
\cite{jordens08}.

The possibility that such a modulation might provide a useful probe was
first suggested by Kollath {\it et al.} \cite{Kollath06}, based on earlier
work with bosonic systems \cite{stoferle04}.  Using a time dependent
Density Matrix Renormalization Group method, it was shown that a peak
existed in the induced double occupation at a frequency $\omega$ which
matched the interaction strength $U$. In this treatment, the
response kernel was approximated to include only changes $\delta J$ in
the hopping operator, neglecting corresponding variation $\delta U$ in
the on-site interactions.  Within this approximation, the authors
emphasized that the measurement was sensitive to near neighbor spin
correlations, and the exchange gap, as well as the charge gap.

This `modulation spectroscopy' has been further explored theoretically
by Huber\cite{huber09} and Sensarma\cite{sensarma09}.  In the former
work, the frequency dependence of the shift in $D$ was studied in the
atomic and two particle limits, and within a slave boson mean field
theory.  The latter work focused on observing local antiferromagnetic
order at the superexchange scale.  As with the earlier study of Kollath,
in both of these papers, the modulation was assumed to couple only to the
kinetic energy.

In this paper, we extend previous work by studying the effect
of both the modulation of the tunneling strength $\delta J$ and of
the on-site interaction strength $\delta U$ due to varying the optical
lattice depth $V$, for the two dimensional repulsive fermionic Hubbard
Hamiltonian. The modulation by $\delta U$ is shown to be quite
significant in the parameter range of interest to current experiments.
We find that the filling of the system plays a very important role in
the response.  Crucially, through the use of Determinant Quantum Monte
Carlo (DQMC) \cite{blankenbecler81} and the maximum entropy method
\cite{gubernatis91,MaxEnt}, we provide results which treat the
electron-electron correlations exactly.

In the low energy limit, two species of repulsively interacting fermions
confined to a periodic optical potential with wavelength $\lambda$ and
amplitude $V(t)$ can be described by the one-band Hubbard model
\cite{Bloch},
\begin{equation}
 \hhat =-J \khat + U \dhat - \mu \Nhat,
\label{Hamiltonian}
\end{equation}
where the hopping or kinetic-energy operator is 
$\khat= \sum_{<ij>,\sigma}[\chat^\dagger_{i\sigma}\chat^{\vphantom{\dagger}}_{j\sigma} + {\rm h.c.}]$,  
$\dhat = \sum_i \nhat_{i\uparrow} \nhat_{i\downarrow}$ is the double occupancy,  
and $\Nhat=  \sum_i \nhat_{i\uparrow}+ \nhat_{i\downarrow}$, the total number of particles,  
with $\chat_{i\sigma}^\dagger$($\chat_{i\sigma}$)
the fermion creation (annihilation) operator, $\sigma=\uparrow$,$\downarrow$ the spin index,
$\nhat_{i\sigma} = \chat^\dagger_{i\sigma}\chat^{\vphantom{\dagger}}_{j\sigma}$, 
and $\mu$ the chemical potential.
The hopping ($J$) and  interaction ($U$) can be expressed
as~\cite{Bloch} 
 $J \approx (4\,/\sqrt{\pi}) \, E_R
 \, v^{3/4}\, {\rm exp}(-2 \sqrt{v} )$ and 
 $U \approx
 4\sqrt{2\pi} \, (a_s/\lambda)\, E_R \, v^{3/4}$,
where $v=V/E_R$ is the ratio of lattice
depth to recoil energy, and $a_s$ is the short ranged $s$-wave
scattering length.

It is clear from these expressions that a small
time-dependent modulation of $V$ changes
both $J$ and $U$.  Writing
$V(t)  = V_0 +\delta V\sin(\omega t)$ and expanding $J$ and $U$ in the limit
$\delta V\ll V_0$ yields $\hhat= \hhat_0 + \delta\hhat\sin(\omega t)$ with 
$\hhat_0$ given by Eq.~(\ref{Hamiltonian}) with $J$ replaced by  $J_0$ and $U$ by $U_0$, and 
$\delta\hhat = -\delta J  \khat+  \delta U \dhat$ with the time-dependent perturbations
\bea
\delta J &=&  J_0 \Big(\frac{3}{4}-\sqrt{\frac{V_0}{E_R}} \Big)\frac{\delta V}{V_0},
\nonumber \\
\delta U &=&\frac{3}{4} U_0 \frac{\delta V}{V_0}.
\eea
For $\delta V>0$, we have $\delta J<0$ and $\delta U>0$ so that
 an increase in the optical lattice amplitude suppresses hopping
and increases the Hubbard repulsion. 
We emphasize that one cannot a priori neglect $\delta J$ or $\delta U$ as they
can be of the same order of magnitude if using the experimental parameters 
as in Ref.~\onlinecite{jordens08}.

Our aim is to understand how such a simultaneous modulation of the
hopping and interaction parameters, as provided by fermions in a
time-dependent optical lattice, probes fermion correlations in the
Hubbard model.  To this end, we study the time dependence of the average
double occupancy $D(t) = \langle \dhat\rangle$.
 Within standard time-dependent perturbation theory,
$D(t)$ satisfies, to linear order,
\be
D(t)  =D(t_0)  -  i\int_{t_0}^t dt' \, \langle 
[\dhat(t),\delta\hhat(t')]\rangle_0 \sin\omega t',
\label{eq:dvst1}
\ee
where $\langle \curO \rangle_0 = Z_0^{-1}\text{Tr} e^{-\beta \Hh_0}\curO$ and
$\curO(t) =  {\rm e}^{i\Hh_0t} \curO {\rm e}^{-i\Hh_0t}$.
Equation~(\ref{eq:dvst1}) can be simplified by rewriting $\delta\hhat$
in terms of $\hhat_0$ as $\delta\hhat =
(\delta J / J_0 ) \, (  \Hh_0 + U_0 [\alpha - 1]
\dhat )$,  
with $\alpha =\Big(1-\frac{4}{3}\sqrt{\frac{V_0}{E_R}}\, \Big)^{-1}$.  
When inserted into Eq.~(\ref{eq:dvst1}), the first term will give a
vanishing contribution, leading to 
\be
D(t) = D(t_0) +  \frac{U_0}{J_0} (\alpha -1) \int_{t_0}^\infty dt' \, \delta J 
\chi_{\rm DD}(t-t')\sin\omega t',
\label{eq:dvst3}
\ee
where 
$\chi_{\rm \mathcal{O} \mathcal{O}}(t-t') = -i\langle
[\curO(t),\curO(t')]\rangle_0 \, \theta(t-t')$.
Formally setting $\alpha = 0$ amounts to neglecting the modulation of the
interaction term.  In contrast,  experimentally, $\alpha$ typically
varies within the range $-0.41 < \alpha < -0.28$.  The simplification
leading to Eq.~(\ref{eq:dvst3}), can be generalized to show that
$\chi_{\rm DD}(t) = \big(J_0/U_0)^2 \chi_{\rm KK}(t)$, 
a fact that we shall use below in our analysis.

Numerically, we calculate the imaginary-time quantity
$\chi_{\rm DD}(\tau)$ from Determinant Quantum Monte Carlo simulations
\cite{blankenbecler81} and analytically extrapolate to the
corresponding imaginary part of the real frequency quantity
$\chi''_{\rm DD}(\omega)$  by inverting
\begin{eqnarray}
&& \chi_{\rm DD}(i\nu_n) = -\frac{1}{\pi}\int^{\infty}_{-\infty} d\omega
 \, \frac{\chi''_{\rm DD}(\omega)}{i\nu_n - \omega},
\label{dd_relation}
\end{eqnarray}
via the Maximum Entropy method
\cite{gubernatis91,MaxEnt}. In Eq.~\ref{dd_relation}
$i \nu_n=2n\pi T$ is the bosonic Matsubara frequency, $T$ is the
temperature, and $\omega$ the real frequency.

\begin{figure}[ht]\vspace{0.62cm}
\includegraphics[width=3.375in, height=2.5in]{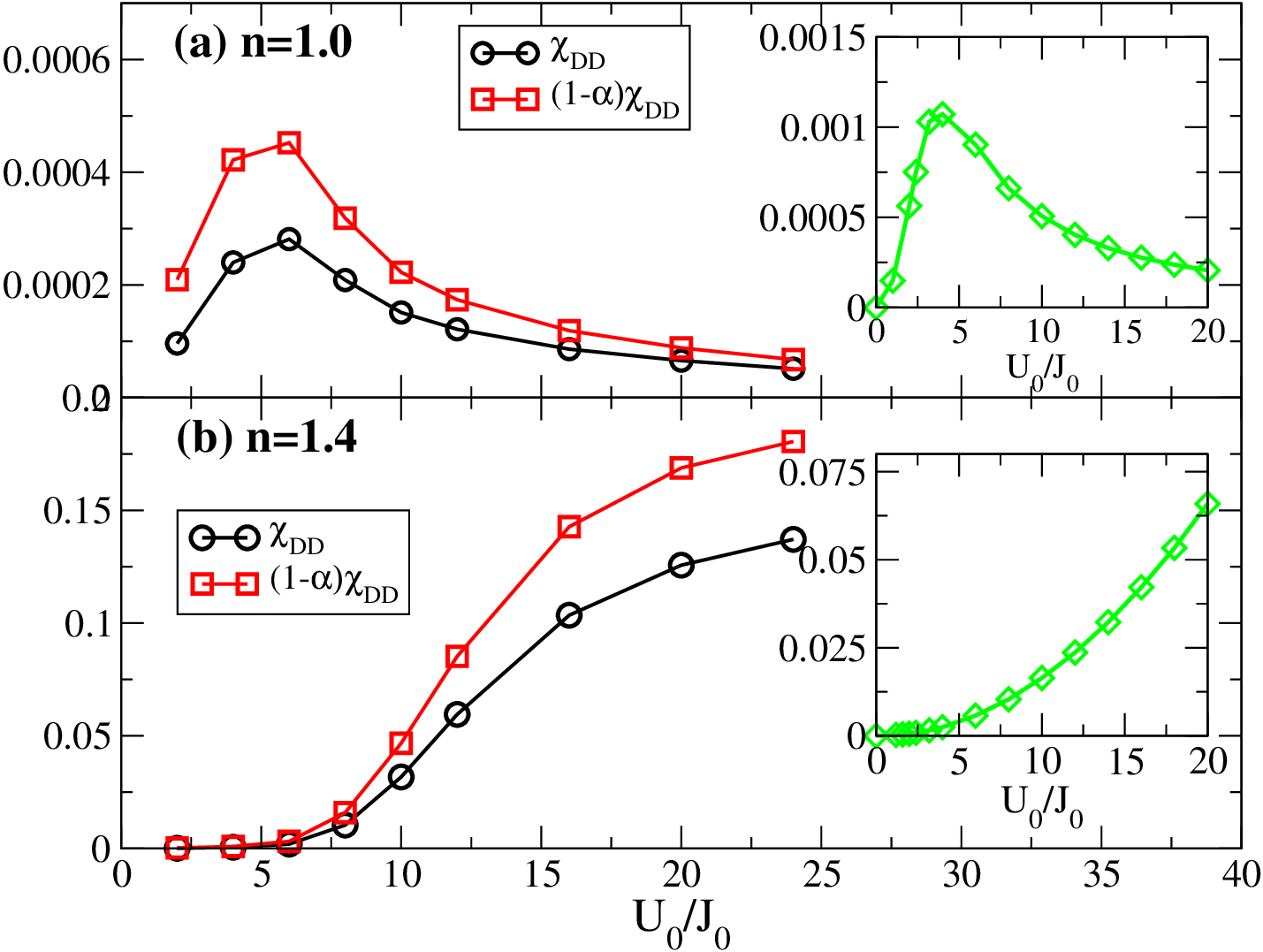}
\caption{(Color online).   
  Top panel (a) shows data for half filling, and panel (b) for a filling of $n=1.4$, for a 
  two-dimensional $4 \times 4$ Hubbard lattice. Red curves (squares) show the quantity 
  $(1-\alpha)\chi_{\rm DD}$, that appears in the linear response of the double occupancy, 
  evaluated at zero Matsubara frequency as a function of $U_0/J_0$. Neglecting the modulation of
  the Hubbard interaction amounts to setting $\alpha =0$, yielding a
  smaller result (black curve, circles). For comparison, the green
  diamonds in the insert in both (a) and (b) are exact results for
  $(1-\alpha)\chi_{\rm DD}$ for a two-site Hubbard model. $\alpha$ is
  determined by assuming $a_s/\lambda = 0.0119$, where $a_s=240 a_0$
  ($a_0$ is Bohr radius) and $\lambda=1,064$ nm (following Ref.~\cite{jordens08}), 
  thus $\alpha$ can be found as a single-valued function of $U_0/J_0$.}
\label{fig_xddT}
\end{figure}

To illustrate the importance of incorporating the modulation of the
interaction parameter $U$, in Fig.~\ref{fig_xddT} we show the
dependence with $U_0/J_0$ of the 
double-occupancy response function $\chi_{\rm DD}(i\nu_n=0)$ (black
curves), for $n=\left\langle n_{i\uparrow} + n_{i\downarrow} \right\rangle = 1.0$ and $n=1.4$ along with this quantity 
multiplied by $(1-\alpha)$ (red curve). Therefore, the black curves is the result from modulating $\delta J$ only, while the red curve also includes the effect of modulating  $\delta U$.  The difference between the curves illustrates that $\delta U$ should not be neglected.
We observe from Fig.~\ref{fig_xddT} that at half-filling
(n=1), the double occupancy response is largest in the
intermediate interaction region and decreases with increasing
$U_0/J_0$.  This is in striking contrast to the behavior at 
$n=1.4$, in which the double occupancy response is
small at weak coupling and saturates at large $U_0/J_0$. 
To confirm our numerical calculation, we analytically
solved the case of a two-site Hubbard model and found qualitatively
similar behavior.  (See green curves in Fig.~\ref{fig_xddT}.)

\begin{figure}[ht]\vspace{0.5cm}
\includegraphics[width=3in, height=2.3in]{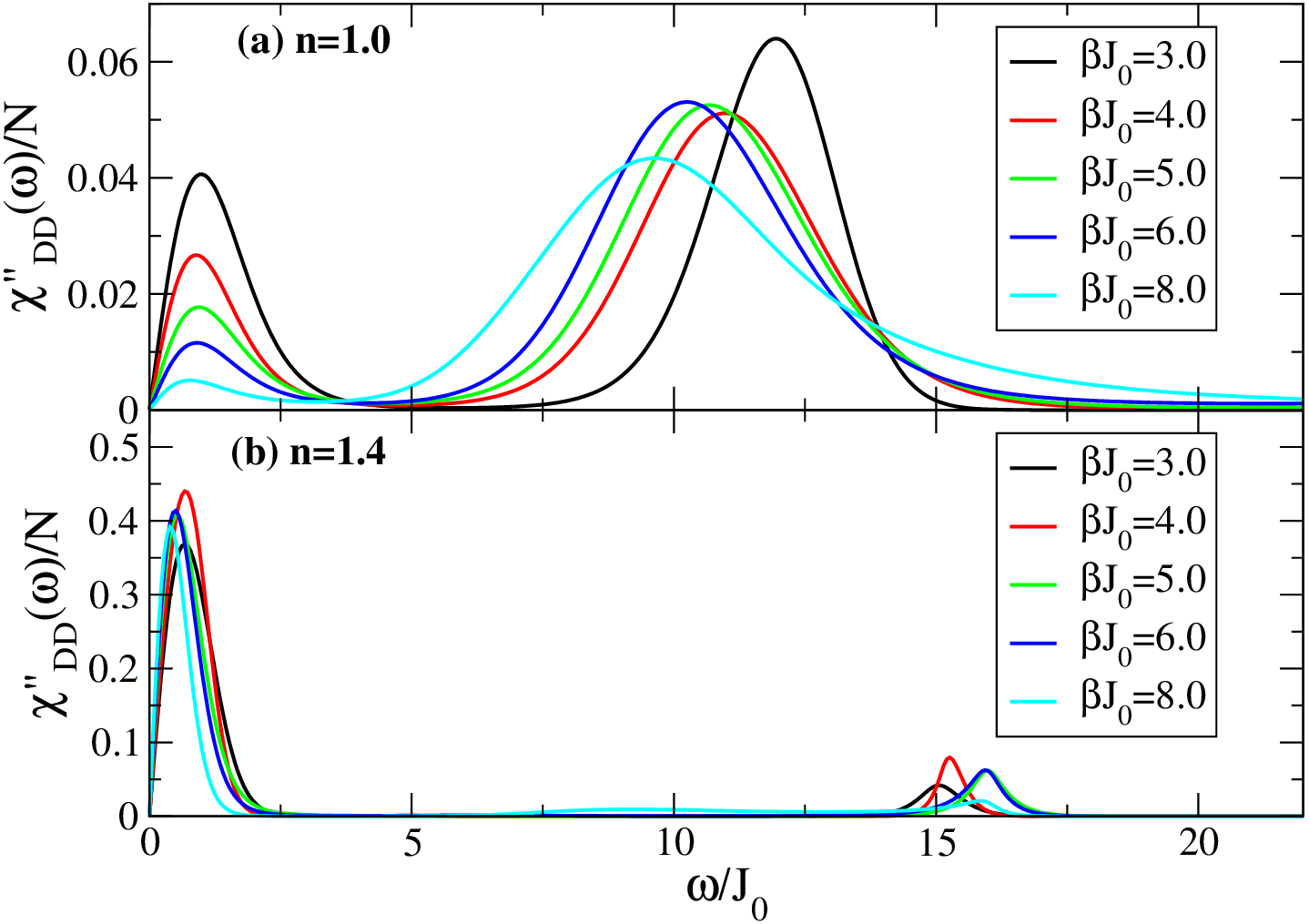}
\caption{(Color online).   
The imaginary component of the double-occupancy susceptibility 
  $\chi''_{\rm DD}(\omega)/N$ for $U_0/J_0=10.0$, a  $4\times 4$ square lattice,
and  various values of inverse temperature ($\beta = 1/T$). 
Panel (a) shows  half-filling $n=1.0$ results, and panel (b) 
a filling of $ n=1.4$. $N=16$ is the system size.}
\label{fig_xddw1}
\end{figure}

We now turn to the full frequency dependent dynamical susceptibility,
which determines the response to the dynamical modulation, showing its
evolution as a function of temperature (expressed in terms of $\beta
J_0 = J_0/\kb T$) in Fig.~\ref{fig_xddw1}.  Panel (a) displays 
results at half-filling, where Mott-insulating physics dominates.
At this filling the low frequency response is strongly suppressed for 
temperatures approaching zero (so that this quasi-peak represents thermally-excited states, not coherent
excitations), with the predominant response occurring at frequencies
close to $U_0$.   This energy scale, corresponding to the Mott gap, is
consistent with recent experimental results~\cite{jordens08} which find a
strong response in the double occupancy when $\omega \sim U_0$. The
presence of the Mott gap also accounts for the much smaller values
of $\chi''$ in the top panels of
Figs.~\ref{fig_xddT} and \ref{fig_xddw1}. Panel (b) shows a filling $ n= 1.4$, 
where an $\omega = 0$ peak remains robust for $T\to
0$.  We attribute this peak to the presence of gapless excitations
reflecting Fermi liquid behavior in this region. The peak at high
$\omega$ represents coherent excitations at the band-gap scale which
should be the distance between the lower and upper Hubbard bands.

\begin{figure}[ht]\vspace{0.5cm}
\includegraphics[width=3.375in, height=2.3in]{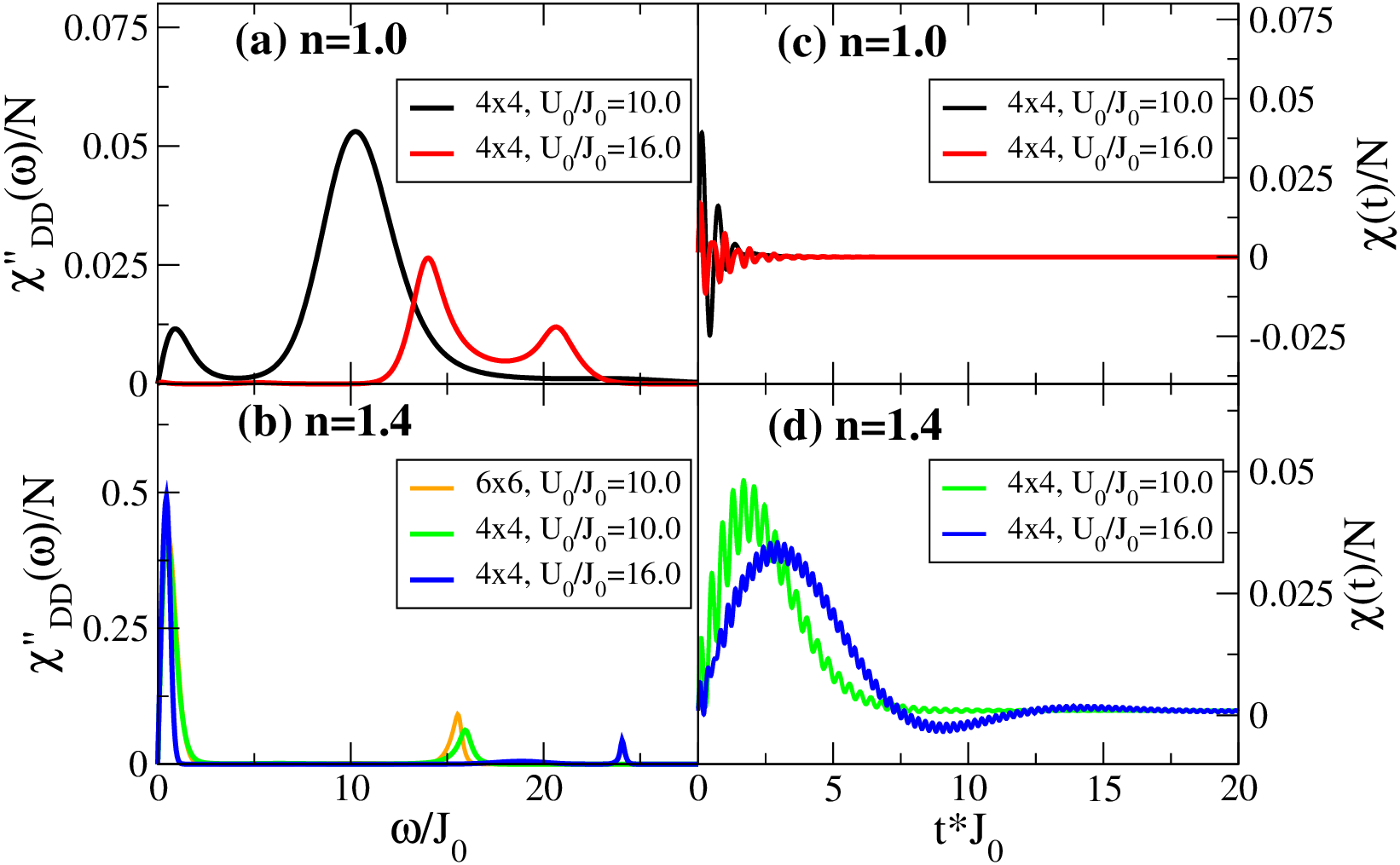}
\caption{(Color online).   
Left column: The imaginary part of the double-occupancy susceptibility
  $\chi''_{\rm DD}(\omega)/N$ for $U_0/J_0=10$ and $16$.
Panel (a) shows  half-filling $n=1.0$ results for a $4\times 4$ lattice, 
$U_0/J_0=10.0$ (black solid curve) and $U_0/J_0=16.0$ (red solid curve). 
Panel (b) shows results for a filling $n=1.4$ and 
for $U_0/J_0=10.0$, $6\times 6$ square lattice (orange curve), 
$U_0/J_0=10.0$, $4\times 4$ (green curve), and $U_0/J_0=16.0$, $4\times 4$ (blue curve).
Right column: The real-time double-occupancy response function $\chi_{\rm DD}(t)$ 
for a $4 \times 4$ square lattice at half filling (panel (c)) for $U_0/J_0=10.0$ 
(black solid curve) and $U_0/J_0=16.0$ (red solid); and for $n = 1.4$ (panel (d)) with 
$U_0/J_0=10.0$ (green curve) and $U_0/J_0=16.0$ (blue curve). All results are 
at a temperature $T/J_0 =2/3$.}
\label{fig_xddw2}
\end{figure}

In Fig.~\ref{fig_xddw2}, we show the interaction dependence of
$\chi_{\rm DD}''(\omega)$. Panel (a) displays the half-filled case where 
the peaks are centered at  $U_0$.  In panel (b), filling $n= 1.4$, 
we include the case of a larger lattice size ($6\times 6$) to show that finite size effects are small. 
These results further verify the important role of filling in the
response to dynamical modulation. Our findings can be
qualitatively reproduced by neglecting vertex corrections in 
$\chi_{\rm KK}$ and expressing the single particle Green's 
function in the Hubbard-I approximation. The latter
corresponds to using a approximate self-energy of the form
\begin{equation}
 \Sigma_\sigma(\omega) \sim \frac{U_0^2 \, n_{\bar{\sigma}}(1-n_{\bar{\sigma}})}{\omega+i\delta}.
\label{eq:Hubbard-I}
\end{equation}
We find that $\chi''_{\rm KK}(\omega)$ (and hence  $\chi''_{\rm DD}(\omega)$) 
possess poles at $\omega  \sim 0, \pm
\sqrt{(\epsilon_{\ark})^2+4 U_0^2 \, n_{\sigma}(1-n_{\sigma})}$, 
where $\epsilon_{\ark}$ is the energy of a non-interacting quasiparticle  
with momentum $\ark$. In the low energy region, there are quasi-elastic peaks at approximately
$\omega \sim 0$. Note that the peak vanishes at $\omega = 0$
because the imaginary part of the real frequency susceptibility is an odd
function $\chi''_{\rm KK}(-\omega) = -\chi''_{\rm KK}(\omega)$.  In the
high energy region, the peaks are located at roughly 
$\displaystyle{\omega \sim U_0 \, + \frac { \epsilon^2_{\ark}}{2 U_0}}$.
Therefore, at half-filling, the peaks are
at $\omega = U_0$ but they sit at higher frequencies away from half filling.

We now turn to the question of how the features in $\chi_{\rm DD}(\omega)$ would be reflected 
in a experimental measurement of the double occupancy, by inserting our results for $\chi_{\rm
DD}(t)$ into Eq.~(\ref{eq:dvst3}).  For this task, we need to obtain
the real part of $\chi_{\rm DD}(\omega)$ via Kramers-Kronig; upon
Fourier transforming we find the real-time dynamical
response functions for the double occupancy to be strikingly different
at half filling and away from half filling, as seen in
panels (c) and (d) of Fig.~\ref{fig_xddw2}. We see that filling $n = 1$
shows a response function that is tightly peaked at $t\to 0$, characterized by a single frequency
scale $\omega \sim U_0$, while at  $n = 1.4$ we see a
broad behavior dominated by the two distinct frequencies associated with
$\omega \sim 0$ and $\displaystyle{\omega \sim U_0 \, + \frac { \epsilon^2_{\ark}}{2 U_0}}$.

\begin{figure}[htb]\vspace{0.5cm}
\includegraphics[width=3.375in, height=2.4in]{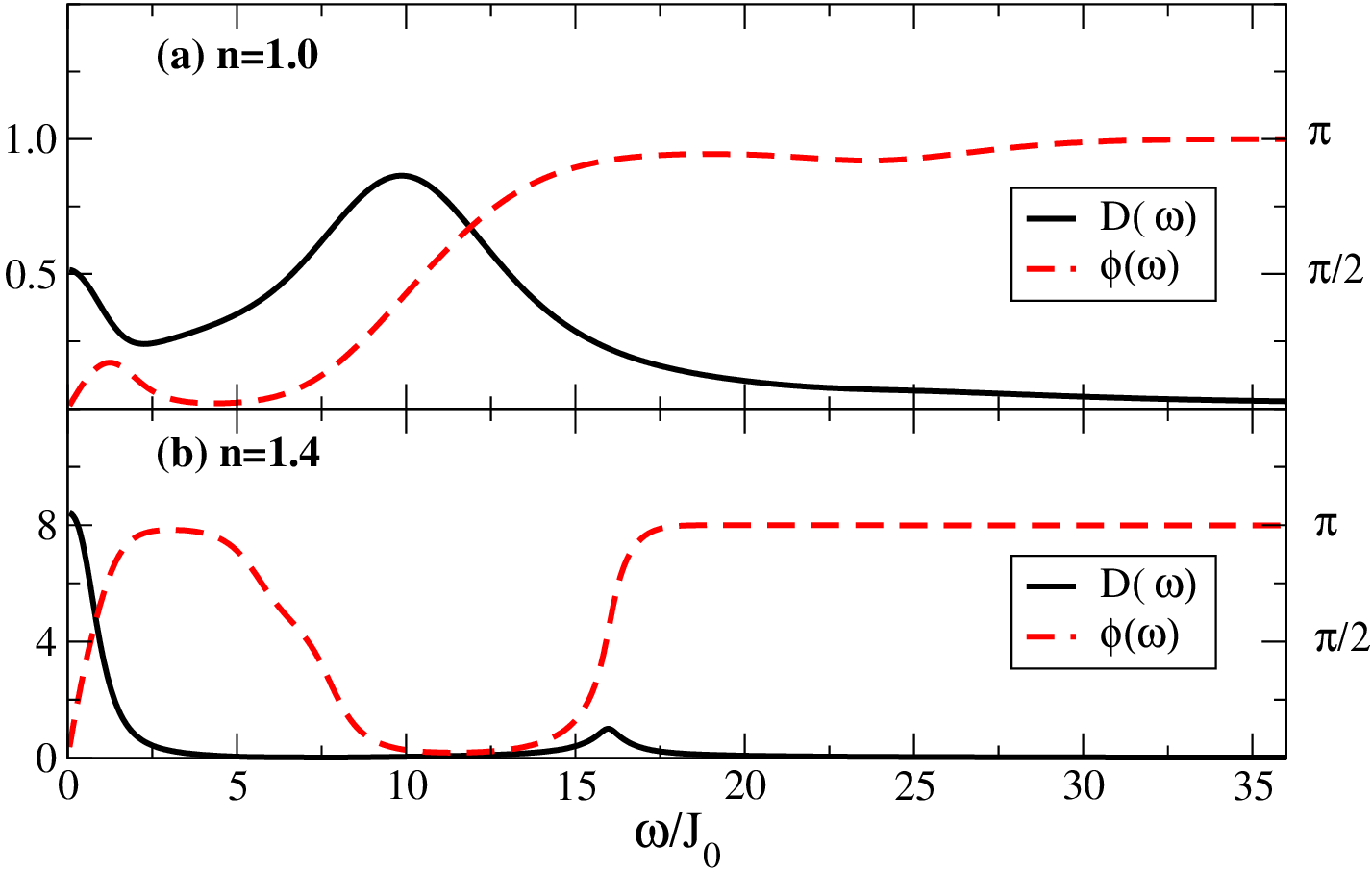} 
\caption{(Color online).   
The frequency dependence of the double occupancy linear response 
for a $4 \times 4$ lattice, interaction strength $U_0/J_0=10.0$, and temperature $T/J_0= 2/3$ .
Panel (a) shows half-filling results; panel (b), $n=1.4$. Solid (black) curves shows the 
amplitude $D(\omega)$ while the dashed (red) curves display the phase shift $\phi(\omega)$ induced 
by the dynamical modulation.}
\label{fig_Dwphi}
\end{figure}

As in standard linear response theory, the real and imaginary parts of
$\chi_{\rm DD}(\omega)$ correspond to the in-phase and out of phase parts
of the response, respectively. Thus, to linear order, an oscillatory driving of the 
optical lattice potential yields an oscillatory response at the
same frequency, but with a phase lag characterized by the ratio of
$\tan \phi(\omega)= \chi''_{\rm DD}(\omega)/\chi'_{\rm DD}(\omega)$. 
This response has recently been observed directly~\cite{Greif}.
We can then write the time-dependent double occupancy as
 \be
\label{eq:dvst}
 D(t)=D(0)+ D(\omega) \sin\lbrack\omega t-\phi(\omega)\rbrack,
\ee
where $D(\omega)=U_0/J_0 (\alpha -1) \delta J |\chi_{\rm DD}(\omega)|$.
We plot $D(\omega)$ and $\phi(\omega)$ in Fig.~\ref{fig_Dwphi} for the case of $U_0/J_0 = 10$.  
We first note that, at low frequency $\omega \to 0$, Eq.~(\ref{eq:dvst}) implies
the time dependence of $D(t)$ to be precisely $\pi$ out of phase with 
$\delta V(t)$. Therefore, an adiabatic increase of the optical lattice amplitude leads to 
a corresponding {\it suppression\/} of the double occupancy.  At higher
$\omega$ these plots show how the time-dependent linear response of the double occupancy probes
the underlying fermion correlations. As we expected, the half filled case shows the strongest
response when the driving frequency $\omega \sim U$, and with a phase that 
is shifted, by $\phi \approx \pi/2$, relative to the imposed modulation.  
At $\langle n\rangle =1.4$, however, the predominant response is for
$\omega = 0$, with phase shift $\phi \approx 0$.

In conclusion, we have investigated the dynamical properties of
fermions in an optical lattice, realized by the Hubbard model subject
to a periodic optical lattice modulation. We show that the modulation of
the on-site interaction cannot be neglected  and that, even at the
level of linear response, the dynamical double occupancy provides a
sensitive probe of fermion correlations. Recent cold-atom
experiments~\cite{Greif} studying the dynamical modulation of the
optical lattice find a linear in time contribution to the double
occupancy, known to emerge at quadratic order in the modulation
parameter $\delta V$~\cite{Kollath06}.  Thus, we expect that our
linear-response results apply at smaller $\delta V/V_0$, or after
subtracting off this $t$-linear contribution to focus on the
oscillatory component.  A future extension of our work will analyze 
the linear and quadratic-order
contributions in detail. In addition, the effects of inhomogeneity due
to trapping effects is an issue for future calculations. 

We gratefully acknowledge discussions with L. Tarruell.  This work is supported by 
NSF OISE-0952300 (ZX, JM, and MJ), DOE CMSN DE-FG02-04ER46129 (ZX), 
DOE SciDAC DE-FC02-06ER25792 (MJ and RTS), ARO W911NF0710576 with funds from
the DARPA OLE Program (RTS), NSF-OCI-0904972 (SC) and by the Louisiana Board of Regents, 
under grant LEQSF (2008-11)-RD-A-10 (DS). Supercomputer support was provided by the 
NSF  TeraGrid under grant number TG-DMR100007.

\end{document}